\documentclass[aip,twocolumn,10pt]{revtex4-1}
\usepackage{graphicx}  
\usepackage{dcolumn}   
\usepackage{bm}        
\usepackage{amssymb}   
\usepackage{amsmath}   
\makeatletter
\newcommand*{\rom}[1]{\expandafter\@slowromancap\romannumeral #1@}
\makeatother

\newcommand{\bra}[1]{\ensuremath{\left\langle#1\right|}}
\newcommand{\ket}[1]{\ensuremath{\left|#1\right\rangle}}

\begin{document}

\title{The Lagrangian Formulation of Strong-Field QED in a Plasma}
\author{Erez Raicher\footnote{E-mail address: erez.raicher@mail.huji.ac.il }}
\affiliation{Racah Institute of Physics, Hebrew University, Jerusalem 91904, Israel }
\affiliation{Department of Applied Physics, Soreq Nuclear Research Center, Yavne 81800, Israel }
\author{Shalom Eliezer}
\affiliation{Department of Applied Physics, Soreq Nuclear Research Center, Yavne 81800, Israel }
\affiliation{Nuclear Fusion Institute, Polytechnic University of Madrid, Madrid, Spain }
\author{Arie Zigler}
\affiliation{Racah Institute of Physics, Hebrew University, Jerusalem 91904, Israel }
\date{\today}

\begin{abstract}
The Lagrangian formulation of the scalar and spinor quantum electrodynamics (QED) in the presence of strong laser fields in a plasma medium is considered. We include the plasma influence in the free Lagrangian analogously to the "Furry picture" and obtain coupled equations of motion for the plasma particles and for the laser propagation. We demonstrate that the strong-field wave (i.e. the laser) satisfies a massive dispersion relation and obtain self-consistently the effective mass of the laser photons. The Lagrangian formulation derived in this paper is the basis for the cross sections calculation of quantum processes  taking place in the presence of a plasma.
 \end{abstract}

\maketitle


\section{ Introduction}
Over the past decades, the strong-field quantum electrodynamics (QED) has attracted a considerable scientific attention (see Refs. 1,2 and references therein). Its non perturbative nature alters its features substantially with respect to the standard QED. For example, it allows for exotic phenomena  such as vacuum pair creation (Schwinger mechanism \cite{schwinger}) and vacuum polarization \cite{euler}. Due to the rapid developement in high intensity laser technology \cite{HIPER, ELI, XCELS}, intensities of $10^{25} - 10^{26}W/cm^2$ are expected to be within reach in the near future, enabling an experimental exploration of this subject.

In the general case, the interaction of a charged particle with an electromagnetic wave is governed by 4 Lorentz invariant parameters \cite{ritus}. Two of which are the classical parameter
\begin{equation}
\xi \equiv \frac{ea}{m}  
\end{equation}
and the quantum parameter
\begin{equation}
\chi \equiv \frac{1}{m} \sqrt{\left ( \frac{du}{ds} \right ) ^2}>1
\end{equation}
where $s$ is the proper time of the particle, $u^{\mu}$ is the proper velocity,  $e$ and $m$ are the charge and the mass of the particle respectively, $A_{\mu}$ is the vector potential and $a$ is its amplitude $a \equiv \sqrt{-A^2}$. Natural units ($\hbar = c = 1$) are used.
The intuitive interpretation of $\xi$ is the number of laser photons participating in the scattering process of the particle and the electromagnetic wave. As a result, if $\xi \gg 1$, i.e. in the strong-field regime, the scattering processes involve many photons absorption.  The quantum parameter $\chi$ is defined as the particle acceleration in units of Schwinger acceleration $a_s \equiv m$. If it is greater than 1, the particle motion in the electromagnetic field has a quantum nature. In practical units, those quantities are given \cite{DiPiazza} by $\xi= 7.5  \sqrt{I_L[10^{20}W/cm^2]}/\omega_L[eV] $  and $\bar{\chi} =I_L[10^{24}W/cm^2] / \omega_L[eV]$ where $\omega_L$ is the energy of the photon laser and the bar symbol denotes a time averaged value.

The additional two parameters are the electromagnetic field invariants 
\begin{equation}
\mathcal{F} \equiv -\frac{1}{4}F_{\mu \nu}F^{\mu \nu}
\end{equation}
 and 
\begin{equation}
\mathcal{G} =\frac{1}{4}\epsilon_{\alpha \beta \mu \nu} F^{\alpha \beta}F^{\mu \nu} 
\end{equation}
Einstein convention is used to summarize over identical greek indices from 0 to 3.	
The electromagnetic field tensor is related to the vector potential by $F_{\mu \nu} \equiv {\partial}_\mu A_{\nu} - {\partial}_\nu A_{\mu} $ and $\epsilon_{\alpha \beta \mu \nu}$ is the Levi Civita tensor. In terms of the electric field $\textbf{E}$ and the magnetic field \textbf{B} it acquires the familiar form
\begin{equation}
\mathcal{F} =- \frac{1}{2}(\textbf{E}^2 - \textbf{B}^2) \\
\end{equation}
\begin{equation}
\mathcal{G} =- \textbf{E} \cdot \textbf{B} 
\end{equation}
 If the electromagnetic wave is propagating in a vacuum, both invariants are zero. However, if a plasma is present then the wave dispersion is
\begin{equation}
k^2 = \omega_L^2-\vec{\textbf{k}}^2 = m_{ph}^2 \neq 0
\label{eq:av dispersion}
\end{equation} 
where $\vec{\textbf{k}}$ is the laser wave vector and $m_{ph}$ is the effective mass of the laser photons. In such a case $\mathcal{G}$ remains zero and $\mathcal{F}  = -a^2 {m_{ph}}^2 /2 $. 

The presence of a strong laser field with intensity corresponding to $\chi >1, \xi \gg 1$ stimulates a mechanism called QED cascade \cite{sokolov, ridgers, nerush2, pukhov}. During the cascade, an electron is accelerated by the laser field, absorbs many laser photons and emits a gamma photon through a quantum process called the nonlinear Compton scattering. The emitted photon reacts with the laser field and decays into an electron positron pair through the nonlinear Breit-Wheeler process. In the midtime between these quantum emissions, the electron motion is approximately classical.
The QED cascade is a testbed to examine our understanding of the fundumental QED dynamics in the presence of strong fields. In addition, a laser-produced QED plasma was considered in the literature in the context of laboratory astrophysics \cite{liang} and as a potential gamma ray source \cite{nerush}. recently, it was suggested that spontaneous cascades impose a limit on the achievable laser intensity \cite{fedotov}.

Since the electron motion is relativistic ($\xi >1$ ) and quantum ($\chi >1$), the appropriate framework for tackling the emission processes mentioned above is QED. However, as a consequence of the enormous number of photon participating in the process, the standard QED is of no use.
The alternative is the nonperturbative QED formalism, also known as "Furry picture" \cite{furry}. In the nonperturbative attitude, the laser background is considered as a classical field and included in the free part of the Lagrangian. As a result, the laser photon propagators are omitted from the Feynman diagram associated with a given process, and their influence is taken into account through the electron propagator. This "dressed" propagator corresponds to the solution of the Dirac equation of a fermion interacting with an electromagnetic field traveling in vacuum (Volkov solution \cite{Volkov, landau}). Employing the Furry picture, the properties of QED in the nonperturbative regime were thoroughly investigated in recent years \cite{QED2, QED3, QED4, finitePulse, finitePulse3}.

Nowadays, the kinetic calculation of QED cascades is based on a Monte Carlo technique describing the quantum processes, integrated with Particle-In-Cell codes taking into account the collective electromagnetic field influence on the classical motion of the electrons \cite{ridgers, sokolov, pukhov}.
The rates embedded in the Monte Carlo routine rely on the assumption that the electromagnetic field invariants vanish. However, in a previous publication \cite{myPaper} we have shown that if a plasma dispersion relation is assumed for the traveling electromagnetic field, the wavefunction depends on the photon effective mass through the parameter $[eam_{ph}/ (k \cdot p)]^2$ where $p_{\mu}$ is the initial momentum of the particle. If the plasma is initially at rest this expression is equivalent to $(\xi m_{ph} / \omega_L)^2 $. For optical lasers and a dense plasma (i.e. $m_{ph} \sim \omega_L$), it becomes larger than 1 if $I_L > 10^{18} W/cm^2$, which is already available experimentally. It should be stressed that as long as $\chi < 1$ the classical approach is adequate and the quantum mechanics is not required. However, as soon as we enter the QED regime, one cannot neglect the photon effective mass in the presence of a dense plasma. In particular, the quantum rates formulas employed in cascades calculations have to be modified. 

The problem of a particle in the presence of a strong electromagnetic field in a medium was investigated by several authors through the recent decades \cite{felber, becker, mendonca}. It was shown that if one assumes a massive-like dispersion relation for the electromagnetic field then the Dirac / Klein-Gordon equation reduces to the Mathieu ordinary differential equation (using standard manipulations). Moreover, by a proper Lorentz transformation, the electromagnetic field transforms into a homogenous rotating electric (for a timelike wave vector, i.e. $k^2>0$) or magnetic field (for a spacelike wave vector, i.e. $k^2 <0$). Consequently, in the timelike case Schwinger pair production takes place whereas in the spacelike case bands structure is formed. Recently, specific subsets of the solutions space were obtained analytically \cite{varro4,varro5, myPaper}.

The dynamic of the Dirac equation in the presence of a homogenous electric field was thoroughly studied in the literature \cite{ruffini, brezin, mocken, fedotov2, hebenstreit} in the context of Schwinger pair creation. One might wonder how can these investigations shed light upon our problem. Various techniques were used to tackle this subject - WKB approximation \cite{brezin}, Bogolyubov transformation \cite{mocken}, quantum Vlasov equation \cite{fedotov2}, Dirac-Wigner phase space calculations \cite{hebenstreit}, etc. These non-perturbative attitudes are appropriate for this purpose but are not optimal for the calculation of QED processes relevant in the context of QED cascades. These are regularly treated by the standard "Furry picture" mentioned above, which we generalize in the present work. In addition, the attitude of Refs. 23-28, which we adopt, might be used as another alternative approach to the problem of vacuum pair production.

The coupling between the quantum dynamics of a single particle (Dirac / Klein-Gordon / Schr\"{o}dinger) and the plasma collective motion was also discussed through the years, in the context of quantum plasmas \cite{shukla_rev}. The main approaches to this issue are the Wigner formalism \cite{wigner} and the quantum fluid model \cite{marklund}. Both were also generelized for relativistic quantum plasmas \cite{hakim, asenjo} and applied to dispersion relation calculation of weak-amplitude plasma waves (linear response). Recently, the dispersion relation of strong transverse electromagnetic waves traveling in relativstic quantum plasmas was found numerically \cite{shukla_D, shukla_KG}.

In this work we construct the Lagrangian formalism of strong-field QED in the presence of a plasma. The motivation is twofold: first, it will allow us in the future to conduct cross section calculation taking the plasma effect into account. Secondly, it will prove that the dispersion relation indeed takes the massive-like form (\ref{eq:av dispersion}), as was phenomenolgically assumed in the previous articles mentioned above \cite{felber, becker, mendonca, varro4, varro5, myPaper}. Moreover, it will provide us with a self consistent equation for the photon effective mass in strong-field waves.

This paper is organized as follows. In section \rom{2}. we construct the Lagrangian formulation of QED in a plasma induced by strong fields (scalars and fermions are treated seperately). In section \rom{3}. the solutions of the free field equations obtained in a previous paper \cite{myPaper} are reviewed and used in order to obtain the current associated with the particles motion. Section \rom{4}. describes the self consistent calculation of the effective photon mass appearing in the dispersion relation. In section \rom{5}. we discuss the implications of our model and conclude this study.

\section{Lagrangian Formalism}

\subsection{Scalars}
The scalar QED Lagrangian is given by
\begin{multline}
\mathcal{L}_{sQED} =\frac{1}{2} \left[ \left(\partial_{\mu} +ieA_{\mu} \right) \Phi^* \right] \left[ \left( \partial^{\mu} - ieA^{\mu} \right) \Phi \right] \\ -\frac{1}{2} M^2 \Phi^* \Phi - \frac{1}{16 \pi}F_{\mu \nu}F^{\mu \nu}
\label{eq:av sLag}
\end{multline}
where $\Phi$ is a charged scalar field, $\Phi^*$ is its complex conjugate and the electromagnetic field strength tensor is
\begin{equation}
F_{\mu \nu} = {\partial}_\nu A_{\mu} - {\partial}_\mu A_{\nu} 
\end{equation}
The Euler-Lagrange equation of motion for the scalar field 
\begin{equation}
\frac{\delta \mathcal{L}}{\delta \Phi^*}=\partial_{\mu} \frac{\delta \mathcal{L}}{\delta ( \partial_{\mu} \Phi^*)}
\end{equation}
yields the familiar Klein-Gordon equation.
\begin{equation}
\left [- {\partial}^2  +M^2 \right] \Phi =  \left[ -e^2A^2 + 2ie A \cdot \partial  \right] \Phi
\label{eq:av KG1}
\end{equation}
where $\partial^2 \equiv \partial^{\mu} \partial_{\mu}$, $A^2 \equiv A^{\mu} A_{\mu}$ and $A \cdot \partial \equiv A^{\mu} \partial_{\mu}$.

According to the Noether theorem, if the Lagrangian is invariant with respect to an infinitisimal transformation $\Phi \rightarrow \Phi + \alpha \Delta \Phi$ then the system admits a conservation law $\partial_{\mu} J^{\mu} = 0$ where the conserved current is
\begin{equation}
J_{\mu} = \frac{\partial \mathcal{L}}{\partial \left( \partial_{\mu} \Phi   \right)} \Delta \Phi
\end{equation}
The conserved current associated with the invariance of the scalar QED Lagrangian under the phase transformation $\Phi \rightarrow \Phi e^{i \alpha} \approx \Phi +i \alpha \Phi $ is
\begin{equation}
J_{\mu} =  \left[  \Phi \left( i\partial_{\mu} - eA_{\mu} \right) \Phi^*   -    \Phi^* \left( i\partial_{\mu} +eA_{\mu} \right)  \Phi \right]
\end{equation}
This current appears as a source in the Euler-Lagrange equation for $A_{\mu}$ 
\begin{equation}
\partial^2 A_{\mu} = -4 \pi  e J_{\mu} 
\label{eq:av Max1}
\end{equation}
Other 4 conserved currents correspond to the translation invariance of the Lagrangian under $x_{\mu} \rightarrow x_{\mu} + d_{\mu}$ (where $d_{\mu}$ is a constant 4-vector)
\begin{equation}
T_{\mu \nu} = \frac{\partial \mathcal{L}}{\partial \left( \partial^{\mu} \Phi   \right)} \partial_{\nu} \Phi - g_{\mu \nu} \mathcal{L}
\end{equation}
$T_{\mu \nu}$ is the energy-momentum stress tensor and $g_{\mu \nu}$ is the metric tensor defined by $g_{00} = 1, g_{11}=g_{22}=g_{33}=-1$ and $g_{\mu \nu} = 0$ for $\mu \neq \nu$. $T_{00}$ is the Hamiltonian density.

In standard QED, the free part of the Lagrangian is
\begin{multline}
\mathcal{L}^0_{free} =\frac{1}{2}\partial_{\mu}\Phi^* \partial^{\mu}\Phi  -  \frac{1}{2} M^2 \Phi^* \Phi -    \frac{1}{16 \pi}F_{\mu \nu}F^{\mu \nu}  
\label{eq:av sLag2}
\end{multline}
while the remaining terms, which couple the scalar and the photon field operators, are considered as a perturbation
\begin{equation}
\mathcal{L}^0_{int}  = A \cdot ( f_1 + A f_2 ) 
\label{eq:av Lint1}
\end{equation}
where
\begin{equation}
f^{\mu}_1 \equiv  \frac{1}{2} ie\left(  \Phi^* \partial^{\mu} \Phi - \Phi \partial^{\mu} \Phi^*   \right) 
\label{eq:av f1_def}
\end{equation}
\begin{equation}
f_2 \equiv \frac{1}{2} e^2 |\Phi|^2
\label{eq:av f2_def}
\end{equation}
However, in the presence of strong electromagnetic fields, $\mathcal{L}^0_{int}$ is no longer a small perturbation. In order to overcome this obstacle, we adopt the "Furry picture" \cite{furry} which is the regular technique in strong-field calculations. Accordingly, we decompose the electromagnetic vector potential into 2 parts
\begin{equation}
A_{\mu} = A^{cl}_{\mu} + A^{Q}_{\mu}
\label{eq:av A_decomp}
\end{equation}
where 
\begin{equation}
A^{cl}_{\mu}   \equiv \bra{\Omega}  A_{\mu}  \ket{\Omega}
\end{equation}
is the expectation value of the vector potential with respect to the ground state of the system $\ket{\Omega}$. The ground state contains the ambient plasma and the laser under consideration and will be defined later on. The quantum part is
\begin{equation}
A^{Q}_{\mu}   \equiv A_{\mu}  -   \bra{\Omega}  A_{\mu}  \ket{\Omega}
\end{equation}
The classical part represents the laser field while the quantum part stands for the emitted radiation.

In the physical scenario of interest, i.e. laser propagation through a plasma, the operators $f_1,f_2$ have a non vanishing expectation value as well. As a result, we generelize Furry attitude and decompose them in an analogous way
\begin{equation}
f^{cl,\mu}_1 = \bra{\Omega}  f^{\mu}_1  \ket{\Omega}
\label{eq:av f1}
\end{equation}
\begin{equation}
f^{Q,\mu}_1 \equiv  f^{\mu}_1 - f^{cl,\mu}_1
\end{equation}
and
\begin{equation}
f^{cl}_2 = \bra{\Omega}  f_2  \ket{\Omega}
\end{equation}
\begin{equation}
f^{Q}_2 \equiv  f_2 - f^{cl}_2
\label{eq:av f2}
\end{equation}
Substituting (\ref{eq:av A_decomp} - \ref{eq:av f2}) into (\ref{eq:av Lint1}) one gets 
\begin{multline}
\mathcal{L}^0_{int} = A^{cl} \cdot f^Q_1 +  f^Q_2 {A^{cl}}^2 +A^Q \cdot f^Q_1 + \\ 2f^{cl}_2 \left(A^{cl} \cdot A^Q  \right) 
+ 2 f^Q_2 \left(  A^{cl}  \cdot A^Q  \right) \\ +  A^Q \cdot f^{cl}_1  +  {A^Q}^2 \left(  f^{cl}_2 +  f^{Q}_2  \right) 
+ A^{cl} \cdot f^{cl}_1 + f^{cl}_2 {A^{cl}}^2
\end{multline}
The last two terms are c-numbers and therefore can be omitted from the Lagrangian. The free equations of motion shall not couple $A^Q$ and $f^Q_1, f^Q_2$. Hence, the following terms 
\begin{multline}
\mathcal{L}_{Furry} = A^{cl} \cdot f^Q_1 + 2\left(   A^{cl} \cdot A^Q \right) f^{cl}_2 \\ + f^Q_2 {A^{cl}}^2 +A^Q \cdot f^{cl}_1 +{A^Q}^2  f^{cl}_2
\end{multline}
can be transferred to the free Lagrangian, which takes the form
\begin{multline}
\mathcal{L}_{free} =\frac{1}{2}\partial_{\mu}\Phi^* \partial^{\mu}\Phi  -  \frac{1}{2} M^2 \Phi^* \Phi  \\ -  \frac{1}{16 \pi}F_{\mu \nu}F^{\mu \nu}   + \mathcal{L}_{Furry}
\label{eq:av free4}
\end{multline}
The interaction part satisifing $\mathcal{L} = \mathcal{L}_{free} + \mathcal{L}_{int}$ is therefore
\begin{equation}
\mathcal{L}_{int} = 2 f^Q_2 \left(  A^{cl}  \cdot A^Q  \right)  +  A^Q f^Q_1  +  {A^Q}^2   f^{Q}_2  
\label{eq:av LintS1}
\end{equation}
The variations are taken with respect to $A^Q_{\mu},  \Phi$. 
The equations of motion corresponding to (\ref{eq:av free4}, \ref{eq:av LintS1}) are
\begin{multline}
\left [- {\partial}^2 + e^2{A^{cl}}^2 - 2ie A^{cl} \cdot \partial +M^2 \right] \Phi =  \\  A^Q \cdot \left[  2ie  \partial  -e^2 \left( A^Q+2A^{cl} \right)   \right] \Phi
\label{eq:av KG2}
\end{multline}
and
\begin{multline}
\frac{1}{4 \pi}\left[\partial^2 A^Q_{\mu} +\partial^2  A^{cl}_{\mu} \right] -2 f^{cl}_2 A^Q_{\mu}   + e J^{cl}_{\mu} =\\ f^{Q,\mu}_1 + 2\left( A^{cl}_{\mu} + A^Q_{\mu} \right) f^Q_2 
\label{eq:av fullMax}
\end{multline}
where $J^{cl}_{\mu} \equiv \bra{\Omega} J_{\mu} \ket{\Omega}$. The right hand side of Eqs. (\ref{eq:av KG2} - \ref{eq:av fullMax}) stems from the interaction term. Taking the expectation value of (\ref{eq:av fullMax}) we arrive at
\begin{equation}
\partial^2  A^{cl}_{\mu}  + 4 \pi e J^{cl}_{\mu} =0
\label{eq:av free2}
\end{equation}
We substract (\ref{eq:av free2}) from (\ref{eq:av fullMax}) and get an equation of motion for $A^Q_{\mu}$
\begin{equation}
\partial^2 A^Q_{\mu} -8 \pi f^{cl}_2 A^Q_{\mu} =4 \pi \left[  f^{Q,\mu}_1 + 2\left( A^{cl}_{\mu} + A^Q_{\mu} \right) f^Q_2 \right]
\label{eq:av Max4}
\end{equation}
In section \rom{4}. we demonstrate that $4 \pi e J^{cl}_{\mu} = m^2_{ph}A^{cl}_{\mu}$. Consequently, Eq. (\ref{eq:av free2}) likewise Eq. (\ref{eq:av Max4}), has a massive-like structure, resulting in the non-zero effective mass of the photons. The emergence of the gauge field mass ia analogous to the Higgs mechanism in particle physics or to the effective photon mass in superconductors.  
For the sake of clarity we write the unpertubed system of equations with which we deal in the rest of the paper.
\begin{equation}
\left [- {\partial}^2 + e^2{A^{cl}}^2 - 2ie A^{cl} \cdot \partial +M^2 \right] \Phi =  0
\label{eq:av KG3}
\end{equation}
\begin{equation}
\partial^2  A^{cl}_{\mu}  + 4 \pi e J^{cl}_{\mu} =0
\label{eq:av free3}
\end{equation}
\begin{equation}
\partial^2 A^Q_{\mu} -8 \pi f^{cl}_2 A^Q_{\mu} =0
\label{eq:av Max5}
\end{equation}

The detailed solution of Eq. (\ref{eq:av KG3}) was the topic of a previous paper \cite{myPaper} and will be reviewd in the next section. The free field solutions enable us to quantize the fields $\Phi$, $A^Q_{\mu}$
\begin{equation}
\Phi =  \int{ \frac{d^3p}{(2 \pi)^{\frac{3}{2}} \sqrt{2p_0}} \left(  c_p \phi_A (p) + h.c \right) }
\label{eq:av phi_series1}
\end{equation}
\begin{equation}
A^Q_{\mu} =  \int{ \frac{d^3k_s}{(2 \pi)^{\frac{3}{2}} \sqrt{2k^0_s}} \left[ \epsilon_{\mu} a_{k_s} e^{-i k_s \cdot x}+ h.c  \right]} 
\label{eq:av A_quant}
\end{equation}
$\phi_A (p)$ is the free scalar wavefunction with momentum $p$. The free wavefunction for a photon with a polariztion vector $\epsilon_{\mu}$  is $\epsilon_{\mu} e^{-i k_s \cdot x}$ where the wave vector $k_s$ satisfies $k^2_s = 8 \pi f^{cl}_2$. The value of $f^{cl}_2$ is calculated in section \rom{4}. The operators $a_{k_s}, c_p$ annihilate photons and scalars respectively, and their Hermitian conjugate are the creation operators. They obey the commutation relations
\begin{equation}
[a^{\dagger}_{k_s},a_{k_s'}] = \delta_{k_s{k_s}'}
\end{equation}
\begin{equation}
[c^{\dagger}_p,c_{p'} ] = \delta_{pp'}
\end{equation}
\\We address a many body problem, therefore the ground state $\ket{0}$ used in the standard quantum field theory is exchanged by
\begin{equation}
\ket{\Omega} = \ket{\int{\frac{d^3p}{(2 \pi)^{\frac{3}{2}}} \sqrt{f(t,\textbf{p})} c_p^{\dagger}(p)} \ket{0}} \ket{\alpha, \hbar \omega_L}
\label{eq:av ground_KG}
\end{equation}
where $\ket{\alpha}$ is a coherent state representing the laser field, t is the time and $f(t,\textbf{p})$ is the distribution function of the scalars.
 In thermal equilibrium the scalars are distributed according to the Bose-Einstein function \cite{shalom}
\begin{equation}
f(\textbf{p}) = \frac{1}{e^{(p_0-\mu)/T}-1}
\end{equation}
where $T$ is the temperature and $\mu$ is the chemical potential.
In the general case, $f(t,\textbf{p})$ is determined by a kinetic equation.

\subsection{ Fermions}
The QED Lagrangian is given by
\begin{equation}
\mathcal{L}_{QED} =\bar{\Psi} \left (i{\not}\partial -e{\not}A -m \right ) \Psi  - \frac{1}{16 \pi}F_{\mu \nu}F^{\mu \nu}
\label{eq:av Lag1}
\end{equation}
where $\Psi$ is the fermion field, $\gamma_{\mu}$ are the Dirac matrices and Feynamn slash notation is adopted, i.e. ${\not}\partial \equiv \gamma^{\mu}\partial_{\mu}$ and ${\not}A \equiv \gamma^{\mu}A_{\mu}$.  
The Euler-Lagrange equation of motion for the fermion field reads
\begin{equation}
\frac{\delta \mathcal{L}}{\delta \bar{\psi}}=\partial_{\mu} \frac{\delta \mathcal{L}}{\delta ( \partial_{\mu} \bar{\psi})}
\end{equation}
and it results in the familiar Dirac equation
\begin{equation}
\left [i {\not}{\partial}  -m \right] \Psi =  e{\not}A \Psi
\label{eq:av Dirac1}
\end{equation}
The Euler-Lagrange equation for $A_{\mu}$ yields
\begin{equation}
\partial^2 A_{\mu} = -4 \pi e J_{\mu}
\label{eq:av Maxwell1}
\end{equation}
where $J_{\mu}$ is the Noether current associated with the invariance of (\ref{eq:av Lag1}) under the phase transformation
\begin{equation}
J_{\mu} = \frac{\partial \mathcal{L}}{\partial \left( \partial_{\mu} \Psi   \right)} \Psi = \bar{\Psi} \gamma_{\mu} \Psi
\end{equation}
The stress energy tensor is defined similiarly to the scalar case by
\begin{equation}
T_{\mu \nu} = \frac{\partial \mathcal{L}}{\partial \left( \partial^{\mu} \Psi   \right)} \partial_{\nu} \Psi - g_{\mu \nu} \mathcal{L}
\end{equation}
In terms of the current $J_{\mu}$, the Lagrangian may be cast in the form
\begin{equation}
\mathcal{L}_{QED} =\bar{\Psi} \left (i{\not}\partial  -m \right ) \Psi  - \frac{1}{16 \pi}F_{\mu \nu}F^{\mu \nu} -e J \cdot A
\label{eq:av Lag2}
\end{equation}
where the last term is the interaction term.

As in the previous section, we decompose the vector potential and the density current into 2 parts
\begin{equation}
A_{\mu} = A^{cl}_{\mu} + A^{Q}_{\mu}
\end{equation}
\begin{equation}
J_{\mu} = J^{cl}_{\mu} + J^{Q}_{\mu}
\end{equation}
Employing these definitions we have
\begin{equation}
J \cdot A = J^{cl} \cdot A^{cl} + J^{cl} \cdot A^{Q}  +  J^{Q} \cdot A^{cl}  +   J^{Q} \cdot A^{Q}
\end{equation}
The first term is a c-number and does not contribute to the equations of motions. The new interaction term, coupling the light and matter fields, is
\begin{equation}
\mathcal{L}_{int} = -  e   A^Q \cdot  J^Q
\label{eq:av LintD1}
\end{equation}
and the remaining terms are absorbed in the free Lagrangian 
\begin{multline}
\mathcal{L}_{free} =\bar{\Psi} \left (i{\not}\partial -m \right ) \Psi - \frac{1}{16 \pi}F_{\mu \nu}F^{\mu \nu} \\  - e  A^Q   \cdot J^{cl} -  eA^{cl}\cdot J^Q
\label{eq:av LfreeD1}
\end{multline}
where the variations are taken with respect to $A^Q_{\mu}, \Psi$. The analogy with the Higgs mechanism discussed in the previous section applies here as well. The equations of motion derived from (\ref{eq:av LintD1}, \ref{eq:av LfreeD1}) are
\begin{equation}
\left [i {\not}{\partial} -e{\not}A^{cl}  -m \right] \Psi = e{\not}A^{Q} \Psi
\label{eq:av Dirac2}
\end{equation}
\begin{equation}
\partial^2 A^Q_{\mu}   = -4 \pi e  \left(   J_{\mu} -J^{cl}_{\mu}  \right)
\label{eq:av Maxwell4}
\end{equation}
\begin{equation}
\partial^2  A^{cl}_{\mu}  + 4 \pi eJ^{cl}_{\mu} = 0
\label{eq:av Maxwell3}
\end{equation}
where the right hand of (\ref{eq:av Dirac2} - \ref{eq:av Maxwell3}) follows from the interaction term. The detailed solution of the unperturbed equation of motion for the fermion (i.e. Eq. (\ref{eq:av Dirac2}) with a vanishing right hand side) was the topic of a previous paper \cite{myPaper} and will be reviewd in the next section.  

The free field solutions enable us to quantize the field $\Psi$ 
\begin{equation}
\Psi =  \int{ \frac{d^3p}{(2 \pi)^{\frac{3}{2}} \sqrt{2p_0}} \left[  b_p \psi_A(p) + h.c \right] }
\label{eq:av Psi_series1}
\end{equation}
where $ b_p$ ($b^{\dagger}_p$) annihilates (creates) an electron with a momentum p and $\psi_A(p)$ is its free wavefunction. The annihilation and creation operators obey the anti commutation relations
\begin{equation}
\{b^{\dagger}_p,b_{p'} \} = \delta_{pp'}
\end{equation}
The quantization of $A^Q_{\mu}$ is the same as in the scalar case (\ref{eq:av A_quant}), excluding the fact that the free field solution are now $\epsilon_{\mu} e^{-i k_f \cdot x}$ where the wave vector $k_f$ satisfies the dispersion relation $k^2_f = 0$.

Let us now explicitly write down the ground state of our many body system 
\begin{equation}
\ket{\Omega} = \ket{ \mathcal{A}   \int{\frac{d^3p}{(2 \pi)^3} \sqrt{f(t,\textbf{p})}b_p^{\dagger}(p)} \ket{0}} \ket{\alpha, \hbar \omega_L}
\label{eq:av ground}
\end{equation}
Since we consider fermions, the wavefunction must be anti symmetrized using the anti symmetrization operator 
\begin{equation}
\mathcal{A}  \equiv  \frac{1}{N!} \sum_{S_N}{(-1)^\pi \hat{P}}
\end{equation}
where N is the particles number, $\hat{P}$ is a permutation belonging to the permutation group $S_N$ and $\pi$ is its parity.
In thermal equilibrium the fermions are distributed according to Fermi-Dirac function \cite{shalom}
\begin{equation}
f(\textbf{p}) = \frac{1}{e^{(p_0-\mu)/T}+1}
\label{eq:av FD}
\end{equation}
where $T$ is the temperature and $\mu$ is the chemical potential.

\section{ Solution of the Free field Equation}

\subsection{ Scalars}
The solution of the Klein-Gordon equation in the presence of a plasma
\begin{equation}
\left [- {\partial}^2 + e^2{A^{cl}}^2 - 2ie A^{cl} \cdot \partial +M^2 \right] \phi_A =  0
\label{eq:av KG10}
\end{equation}
was derived by the authors in a previous paper \cite{myPaper}. We briefly review the final results and use them in order to calculate the current $j_{\mu}$.

We consider a circular polarized laser taking the form
\begin{equation}
A^{cl}(\phi) = a(\phi) \left ( \epsilon e^{i\phi} +  {\epsilon}^* e^{-i\phi} \right )
\label{eq:av Apot}
\end{equation}
where $a(\phi)$ is a slow envelope going to $0$ at $\phi \rightarrow \pm \infty$ and the polarization vector is
\begin{equation}
\epsilon =( e_{1}-ie_{2} ) / 2 
\end{equation}
where
the unit vectors are $e_{1} = (0,1,0,0)$ and $e_{2} = (0,0,1,0)$.
The polarization vector obeys ${\epsilon}^2 = {\epsilon^*}^2 = 0$ as well as $\epsilon \cdot  \epsilon^*= - \frac{1}{2}$. It can be easily verified that $A^2 = -a^2 $.
The most general dispersion relation is (\ref{eq:av dispersion}) where $m_{ph}$ is unknown. The photon effective mass will be found self-consistently in section \rom{4} using the equation of motion of the classical vector potential (\ref{eq:av free2}).
Since in our model the laser photons aquire mass, the gauge invariance is not satisfied. However, one can explicitly show for $A^{cl}_{\mu}, k_{\mu}$ given above that $k \cdot A = 0$.

The wavefunction obeying (\ref{eq:av KG3}) is
\begin{equation}
\phi_A = D_p e^{-i(p-\nu k) \cdot x} 
\label{eq:av KG_f}
\end{equation}
where $D_p$ is a normalization constant and $\nu$ is defined by
\begin{equation}
\nu \equiv \frac{ k \cdot p  }{ {m_{ph}}^2} \left ( 1 -  \sqrt{1 + \left (  \frac{ ea {m_{ph}} }{ k \cdot p } \right )^2 }  \right )
\label{eq:av nu_D}
\end{equation}
$p_{\mu}$ is the 4-momentum of the scalar where the electromagnetic vector potential vanishes, i.e. $\phi \rightarrow \pm \infty$. It should be mentioned that for the sake of simplicity we have assumed $p \cdot A^{cl} = 0$. 

The current density associated with the particle is
\begin{equation}
j_{\mu} =  \left[  \phi_A^* \left( i\partial_{\mu} -eA^{cl}_{\mu} \right)  \phi_A - \phi_A \left( i\partial_{\mu} + eA^{cl}_{\mu} \right) \phi_A^* \right]
\label{eq:av KGcurrent2}
\end{equation}
Substituting (\ref{eq:av KG_f}) into (\ref{eq:av KGcurrent2}) we obtain
\begin{equation}
j_{\mu} = 2 |D_p|^2 \left ( p_{\mu} - eA^{cl}_{\mu} - \nu k_{\mu} \right )
\end{equation}	
The 0 component of $j_{\mu}$ is the charge density, i.e.
\begin{equation}
n = \int{\frac{d^3p}{(2 \pi)^3 2 p_0}f(t,\textbf{p}) \cdot 2 |D_p|^2 \left ( p_{0} - \nu k_{0} \right ) }
\label{eq:av density1}
\end{equation}
where $n$ is the scalar particles number density. In order to preserve the total neutrality, we adopt the simpler possible approximation, i.e. a static positively charged background (e.g. ions) is assumed. Including the ions motion is straightforward but not essential for the purpose of this study. This assumption is applied also in the fermion case discussed in the next subsection.

In addition, the distribution function is related to the number density through
\begin{equation}
n = \int{\frac{d^3p}{(2 \pi)^3}f(t,\textbf{p}) }
\label{eq:av dens2}
\end{equation}
We substitute Eq. (\ref{eq:av dens2}) into (\ref{eq:av density1}) and obtain the normalization coefficient (without solving the integral)
\begin{equation}
D_p = \sqrt{\frac{p_0}{ p_{0} - \nu k_{0}}}
\end{equation}
The final expression for the current is
\begin{equation}
j_{\mu} = 2 \frac{p_0}{ p_{0} - \nu k_{0}} \left ( p_{\mu} - eA^{cl}_{\mu} - \nu k_{\mu} \right )
\label{eq:av KGcurrent3}
\end{equation}

\subsection{ Fermions}
Through detailed derivation \cite{myPaper} we have shown that the solution of the Dirac equation in the presence of a classical field with a massive-like dispersion 
\begin{equation}
\left [i {\not}{\partial} -e{\not}A^{cl}  -m \right] \psi_A = 0
\label{eq:av Dirac5}
\end{equation}
is given by the expression
\begin{multline}
\psi_A = \Bigl [  1- \frac{{\not}k}{2m} \left ( \nu - \frac{e^2a^2}{\rho_1} \right ) - \frac{e}{\rho_1} {\not}k  {\not}{A^{cl}}  \\ - \frac{e}{4m} \left ( 1 + \frac{2 (k \cdot p)}{\rho_1} \right ) {\not}A^{cl}      -  \frac{ie {m_{ph}}^2}{2m \rho_1} \frac{d{\not}A^{cl}}{d \phi}      \Bigr ]  e^{-i(p-\nu k) \cdot x} u_p
\label{eq:av psi_f1}
\end{multline}
where
\begin{equation}
{{\rho_{1}}} \equiv  -2(k \cdot p) \sqrt{ 1+ \left ( \frac{ea m_{ph}}{k \cdot p}   \right)^2 }
\end{equation}
the vector potential and $\nu$ are defined in (\ref{eq:av Apot}, \ref{eq:av nu_D}) and $u_p$ is the Dirac spinor.
The current density of a Dirac particle is given by
\begin{equation}
j_{\mu} = \bar{\psi} \gamma_{\mu} \psi
\label{eq:av Dcurrent}
\end{equation}

In order to avoid unnecessary complication in the current calculation, we would like to prove that for intense lasers conditions the term proportional to $d{\not}A^{cl}/d \phi$ may be neglected. For this purpose we have to show that
\begin{equation}
\left|1+\frac{2(k \cdot p)}{\rho_1} \right| \gg  \left| 2\frac{m^2_{ph}} {\rho_1} \right|
\end{equation}
or equivalently
\begin{equation}
2(k \cdot p) \left( \sqrt{ 1+ \left ( \frac{ea m_{ph}}{k \cdot p}   \right)^2 } -1  \right)  \gg 2m^2_{ph}
\label{eq:av approx}
\end{equation}
Let us examine the left hand side of (\ref{eq:av approx}). If $m^2_{ph} <  [(k \cdot p) /  ea]^2$, than 
\begin{equation}
2(k \cdot p) \left( \sqrt{ 1+ \left ( \frac{ea m_{ph}}{k \cdot p}   \right)^2 } -1  \right)  \approx \frac{(eam_{ph} )^2}{k \cdot p}
\end{equation}
resulting in the condition
\begin{equation}
(ea)^2 \gg 2 k \cdot p
\label{eq:av approx1}
\end{equation}
On the other hand, if $m^2_{ph} >  [(k \cdot p) /  ea]^2$ then the left hand side of (\ref{eq:av approx}) is
\begin{equation}
2(k \cdot p) \left( \sqrt{ 1+ \left ( \frac{ea m_{ph}}{k \cdot p}   \right)^2 } -1  \right)  \approx 2eam_{ph}
\end{equation}
leading to 
\begin{equation}
ea \gg m_{ph}
\label{eq:av approx2}
\end{equation}
For optical intense lasers typical numbers are $ea/m >1$ and $ (k \cdot p)/m^2 > 10^{-5}$ . The photon effective mass is unknown yet, but its upper bound is the laser frequency in the reference frame where the plasma is initially at rest ($\omega_L/m \sim 10^{-5}$). As a result, $m_{ph}/m < 10^{-5}$, so that both (\ref{eq:av approx1}) and (\ref{eq:av approx2}) hold, justifying the neglection of $d{\not}A^{cl}/d \phi$.
Thus, the wavefunction reads
\begin{multline}
\psi_A = \Bigl [  1+ c_k{\not}k + c_A {\not}A^{cl} + c_{kA} {\not}k {\not}A^{cl}         \Bigr ]  e^{-i(p-\nu k) \cdot x}u_p
\label{eq:av psi_f2}
\end{multline} 
where we have defined
\begin{equation}
c_A \equiv -\frac{e}{4m} \left(  1 + \frac{2(k \cdot p)}{\rho_{1}}   \right) 
\end{equation}
\begin{equation}
c_k \equiv -\frac{1}{2m} \left( \nu - \frac{e^2a^2}{\rho_{1}} \right)
\end{equation}
\begin{equation}
c_{kA} \equiv  - \frac{e}{\rho_{1}}
\end{equation}
In order to evaluate the current density we sustitute the wavefunction (\ref{eq:av psi_f2}) into Eq. (\ref{eq:av Dcurrent}). The calculation appears in appendix A, and the final expression takes the form
\begin{equation}
j_{\mu} =  \bar{u}_p u_p \left(  p_{\mu} j^p -   e A_{\mu}^{cl}  j^A     -   k_{\mu} j^k   \right)
\label{eq:av Dcurrent3}
\end{equation}
where
\begin{equation}
j^A  \equiv -\frac{2}{e} \left[  \frac{(k \cdot p)}{m} \left(c_Ac_k - c_{kA} \right)  + c_A - c_k c_{kA}m_{ph}^2  \right]
\end{equation}
\begin{equation}
j^p  \equiv  \frac{ 1+a^2(c_A^2-m_{ph}^2c_{kA}^2) - c_k^2 m_{ph}^2 }{m}
\end{equation}
\begin{equation}
j^k  \equiv  - 2 \left[  \frac{(k \cdot p)}{m} \left( c_k^2  +  a^2 c_{kA}^2 \right)  - a^2c_A c_{kA} + c_k  \right]
\end{equation}
In appendix B we prove two algebraic identities, $j^p = j^A$ as well as $j^k = \nu j^p$. Hence, Eq. (\ref{eq:av Dcurrent3}) reads
\begin{equation}
j_{\mu} =  \bar{u}_p u_p  j^p \left(  p_{\mu}  -   e A_{\mu}^{cl}      -  \nu k_{\mu}    \right)
\label{eq:av Jdirac}
\end{equation}
The normalization of the Dirac spinors is determined according to  
\begin{equation}
n = \int{\frac{d^3p}{(2 \pi)^3 2 p_0}f(t,\textbf{p}) j_{0}}
\label{eq:av density2}
\end{equation}
Substituting Eq. (\ref{eq:av dens2}) into (\ref{eq:av density2}) we find that the normalization condition is $j_0 = 2p_0$, implying
\begin{equation}
\bar{u}_p u_p =   \frac{2p_0} {j^p \left(  p_0  -    \nu k_{0} \right)  }
\label{eq:av upup}
\end{equation}
The final expression for the density current reads
\begin{equation}
j_{\mu} = 2 \frac{p_0}{ p_{0} - \nu k_{0}} \left ( p_{\mu} - eA^{cl}_{\mu} - \nu k_{\mu} \right )
\end{equation}
which is identical to (\ref{eq:av KGcurrent3}), leading to the conclusion that the spin does not influence the density current.

\section{ Photon Effective Mass and the Dispersion Relation}

\subsection{ Scalars}

Having achieved the current of a single particle obeying the free field equation, we are able to find the classical current in the ground state. Employing Eq. (\ref{eq:av phi_series1}, \ref{eq:av ground_KG}) we get
\begin{equation}
J^{cl}_{\mu} = \bra{\Omega}  J_{\mu} \ket{\Omega} = \int{\frac{d^3 p}{(2 \pi)^3 2p_0} f(t,\textbf{p}) j_{\mu}(p)}
\label{eq:av Current0}
\end{equation}
Substituting the expression for the current (\ref{eq:av KGcurrent3}) into (\ref{eq:av Current0}) we have
\begin{equation}
J^{cl}_{\mu} =   \int{\frac{d^3 p}{(2 \pi)^3 2p_0} f(t,\textbf{p})}  2 |D_p|^2 \left ( p_{\mu} - eA^{cl}_{\mu} - \nu k_{\mu} \right )
\label{eq:av Current1}
\end{equation}
In the general case, $f(t,\textbf{p})$ is determined by solving a kinetic equation for the many body system. The initial condition is that of a cold boson gas, i.e. $f(0,\textbf{p}) = \delta^3(\textbf{p})$. As long as $\bar{\chi}<1$ (or equivalently $I_L < 10^{24}W/cm^2$ for $\omega_L = 1eV$), the absorption and emission processes are negligible, so that the quantity $p_{\mu}$ of a particle is conserved (we stress that $p_{\mu}$ is not the instantaneous momentum of the particle but its asymptotic one). Therefore, the distribution function does not change in time $f(t,\textbf{p}) = f(0,\textbf{p})$. As a result we are able to solve the integral appearing in (\ref{eq:av Current1}), and the expression of the current simplifies to
\begin{equation}
J^{cl}_{\mu}  =   \frac{ n_e}{ p_{0} - \nu k_{0}}  \left ( p_{\mu} - eA^{cl}_{\mu} - \nu k_{\mu} \right )
\label{eq:av Current3}
\end{equation}
If $I_L > 10^{24}W/cm^2$ and a cascade takes place, then a more complicated treatment is required, involving both kinetics and quantum rates calculation. Such a scenario will be discussed in the next section.
The substitution of the current (\ref{eq:av Current3}) in the classical Maxwell Eq. (\ref{eq:av free3}) yields the dispersion relation
\begin{equation}
{m_{ph}}^2  =   \frac{4 \pi e^2 n_e}{ p_{0} - \nu k_{0}} 
\label{eq:av dispersion2}
\end{equation}
as well as $J^{cl}_{0} = J^{cl}_{3} = 0$.
In order to compare the dispersion relation with the classical Akhiezer-Polovin formula \cite{polovin}, we transform to a frame of reference in which the particles motion is transverse, i.e $\textbf{p}' = \nu \textbf{k}'$. In this frame $J^{cl}_{3}$ is identicaly zero. $J^{cl}_0$ vanishes due to the neutrality assumption mentioned in the previous section.
We define
\begin{equation}
 \gamma_s M   \equiv    p_{0}' - \nu k_{0}' 
\label{eq:av gammA1}
\end{equation}
 In this reference frame $p_0' = \sqrt{M^2 + \textbf{p}'^2}$. We take the square of (\ref{eq:av gammA1}) and obtain
\begin{equation}
M^2 + \nu^2 \textbf{k}'^2 = \gamma_s^2 M^2 + \nu^2 k_0'^2 +2\nu k_0' \gamma_s M
\label{eq:av gammA2}
\end{equation}
Since the formula of $\nu$ involves scalar variables only, we would like to express the other terms in (\ref{eq:av gammA2}) also in a Lorentz invariant form.
\begin{equation}
2\nu k_0' \gamma_s M = 2\nu k_0' (p_0' - \nu k_0') = 2 \nu ( k \cdot p - \nu m_{ph}^2)
\label{eq:av nu_k0}
\end{equation}
where we have used the relation $k_0'^2 = m_{ph}^2 + \textbf{k}'^2$ and $k \cdot p = k_0'p_0' - \nu \textbf{k}'^2$. Substitution of Eq. (\ref{eq:av nu_k0}) into (\ref{eq:av gammA2}) yields
\begin{equation}
\gamma_s^2 M^2  = M^2 +\nu^2 m_{ph}^2 - 2 \nu (k \cdot p)
\end{equation}
By definition, $\nu$ satisfies the identity
\begin{equation}
-\nu^2 m_{ph}^2 + 2\nu M( k \cdot p) + e^2a^2 = 0
\end{equation}
Finally, $\gamma_s$ is given by
\begin{equation}
\gamma_s = \sqrt{1+\left( \frac{ea}{M} \right)^2}
\end{equation}
Therefore, the dispersion relation (\ref{eq:av dispersion2}) reads
\begin{equation}
m_{ph}^2 = \frac{4\pi e^2 n_e'}{\gamma_s M}
\label{eq:av akhiezer}
\end{equation}
where $n_e'$ is the electron density in the reference frame in which the plasma motion is transverse (i.e. $j_z = 0$) . Eq. (\ref{eq:av akhiezer}) is exactly identical to the familiar Akhiezer-Polovin expression for the plasma frequency of a relativistic classical plasma \cite{polovin}.

The equations of motion derived in section \rom{2} contain the expectation value of two additional operators, $f_2$ and $f^{\mu}_1$.
$f_2$ determines the effective mass of the photons associated with the quantum vector potential $A^Q_{\mu}$. According to Eq. (\ref{eq:av Max5}), the effective mass is $m^Q_{ph} = \sqrt{8\pi f^{cl}_2}$. Let us write down $f^{cl}_2$ explicitly. For this purpose, we substitute Eq. (\ref{eq:av phi_series1}) into (\ref{eq:av f2_def}).
\begin{equation}
f^{cl}_2  = e^2 \int{\frac{d^3 p}{(2 \pi)^3 2p_0} f(t,\textbf{p})  |D_p|^2}
\end{equation}
The integral is solved similiarly to (\ref{eq:av Current1}) and the effective mass is given by
\begin{equation}
{m^Q_{ph}}^2  =   \frac{4 \pi e^2 n_e}{ p_{0} - \nu k_{0}} 
\end{equation}
This equation is identical to (\ref{eq:av dispersion2}) and therefore $A^{cl}_{\mu}, A^Q_{\mu}$ have the same dispersion relation.

Finally, we calculate the expectation value of $f^{\mu}_1$ by substituting Eq. (\ref{eq:av phi_series1}) into (\ref{eq:av f1_def}).
\begin{equation}
f^{\mu,cl}_1  =  \int{\frac{d^3 p}{(2 \pi)^3 2p_0} f(t,\textbf{p})  |D_p|^2} \left(  p_{\mu} - \nu k_{\mu} \right) 
\end{equation}
We have previously seen that $J_0 = J_3 = 0$, or equivalently $p_{\mu}=\nu k_{\mu}$. As a result, the expectation value $f^{\mu,cl}_1$ vanishes.

\subsection{ Fermions}
Utilizing the same procedure as for the scalars, we achieve the following form for the classical density current
\begin{equation}
J^{cl}_{\mu} = \int{\frac{d^3 p}{(2 \pi)^3 2p_0} f(t,\textbf{p}) j_{\mu}}
\label{eq:av Maxwell8}
\end{equation}
Substituting the expression for $j_{\mu}$ (\ref{eq:av Dcurrent3}) into (\ref{eq:av Maxwell8}) we have
\begin{equation}
J^{cl}_{\mu} =   \int{\frac{d^3 p}{(2 \pi)^3 2p_0} f(t,\textbf{p})}  2 |D_p|^2 \left ( p_{\mu} - eA^{cl}_{\mu} - \nu k_{\mu} \right )
\label{eq:av Maxwell9}
\end{equation}
As in the scalar case, we assume that the electrons are a cold gas in order to avoid numerical calulation of $f(t,\textbf{p})$.
Thus, the Fermi-Dirac distribution (\ref{eq:av FD}) reduces to $f(t,\textbf{p}) = [ \theta(p_x)  - \theta(p_x - p_F ][ \theta(p_y) - \theta(p_y - p_F) ][\theta(p_z) -  \theta(p_z - p_F) ]$ where $p_F \equiv \left( 3 \pi^2 n  \right)^{\frac{1}{3}}$ is the Fermi momentum and $\theta$ is the Heaviside function. Since $p_F \ll m$, the integrand of (\ref{eq:av Maxwell9}) is approximately constant in the integration region, leading to
\begin{equation}
J^{cl}_{\mu}  =   \frac{ n_e}{ p_{0} - \nu k_{0}}  \left ( p_{\mu} - eA^{cl}_{\mu} - \nu k_{\mu} \right )
\end{equation}
From the Maxwell equation (\ref{eq:av Maxwell3}) we deduce that $J^{cl}_0=J^{cl}_3 = 0$ (as explained above) and that the photon effective mass is
\begin{equation}
{m_{ph}}^2  =    \frac{4 \pi e^2  n_e}{ p_0   -   \nu k_{0} }  
\label{eq:av disp4}
\end{equation}
Notice that Eq. (\ref{eq:av disp4}) is identical to the scalar case given in Eq. (\ref{eq:av dispersion2}). The transformation to the plasma reference frame is the same as in (\ref{eq:av gammA1} - \ref{eq:av akhiezer}), and consequently the dispersion relation reads
\begin{equation}
m_{ph}^2 = \frac{4\pi e^2 n_e'}{\gamma_f m}
\end{equation}
where 
\begin{equation}
\gamma_f = \sqrt{1+\left( \frac{ea}{m} \right)^2}
\label{eq:av disp5}
\end{equation}
The equivalence between (\ref{eq:av disp5}) and (\ref{eq:av akhiezer}) implies that in the case discussed above the spin does not play any role in the plasma dispersion relation. It is not surprising since we have demonstrated in section \rom{3} that the current density is identical for a scalar and for a fermion. Seeing that this current density is the source of the Maxwell equation and therefore determines the effective mass, the agreement between the fermion and scalar dispersion relations is expected.
This result agrees with the finding presented in Ref. 40, where quantum plasma was shown, by numerical calculation, to exhibit the classical Akhiezer-Polovin dispersion relation unless the density is extremely high (corresponding to $ m_{ph} \sim m$).

\section{ Discussion and Conclusions}
In this contribution we have addressed the Lagrangian formulation of the quantum electrodynamics (QED) and the scalar QED in the presence of strong fields (e.g. intense laser) in a plasma medium. For this purpose, we have decomposed the current operator in the Lagrangian into classical and quantum terms analogously to the "Furry picture" of the electromagnetic field. Utilizing this Lagrangian, coupled self-consistent equations of motion for the plasma particles and for the laser beam propagation are derived. The equation of motion for the plasma particles is the Dirac / Klein-Gordon equation in the presence of an electromagnetic wave with a massive-like dispersion relation. Unlike previous publications \cite{felber, becker, varro4, varro5, mendonca, myPaper} dealing with the quantum dynamics of a particle in the presence of electromagnetic field in a medium, the massive-like dispersion was not inserted phenomenologically into the Dirac / Klein-Gordon equation. Its emergence stems naturally from the Euler-Lagrange equation of motion. To be specific, the photons acquire effective mass due to the nonvanishing expectation value of the current operator, similarly to the way the nonvanishing expectation value of the Higgs field endows elementary particles with their mass.

The dispersion relation is found due to the equation describing the laser propagation and depends on the vacuum expectation value of the electric current associated with the plasma. This value is determined by the current of a single particle and by the particles distribution in the phase space. Explicit expression for the single particle current was obtained using the solution of Dirac / Klein-Gordon equation reviewed in section \rom{3}. The distribution function is the input to our model. Now we shell briefly discuss several possibilities.

 1. The plasma is in a thermal equilibrium. As a result, the distribution function is known (Fermi-Dirac for fermions and Bose Einstein for scalars) and the effective mass of the laser photons is given by the numerical integration of Eq. (\ref{eq:av Maxwell9}). If the plasma is cold ($T = 0$) then the integral solution is straightforward and the effective mass can be calculated analytically. We have demonstrated that in this case the photon effective mass is not affected by the spin, i.e. it is identical for a fermion and for a scalar. 
Note that the scenario of a laser interacting with a counter propogating beam (as in Ref. 45), is equivalent to the cold plasma (in the beam rest frame).

2. If a thermal equilibrium is not established, a kinetic model is required (for example, a Particle In Cell code). In principle, the kinteic model has to be coupled to our formalism, since it must be supplied with cross sections for the quantum processes which shift the particles in the phase space. On the other hand, the wavefunction, and hence also the cross sections, depend on $\xi, \chi, \mathcal{F}$ which are determined by the laser propogation and the particles distribution. The relation between the field invariant $\mathcal{F}$ and the photon effective mass appearing in our model is $m_{ph}= -\frac{\mathcal{F}}{2a^2}$, as was shown in the introduction. 
Nowadays PIC modeling overcome this difficulty in the following way.
The quantites $\xi, \chi$ are evaluated at each time step for each particle according to the electromagnetic field at its position. Then, closed formulas of the quantum processes rates are utilized in order to calculate the probabilty of a particle to be involved in a certain process. Currently, the influence of the field invariant $\mathcal{F}$ is neglected. This neglection is not justified since the particle wavefunction deviates from Volkov solution significantly in the expected conditions in a future laser produced QED cascade \cite{myPaper}. We suggest to generalize this attitude in order to take into account the non vanishing photon effective mass. 
Hence, the input of the cross section section calculation will include also $\mathcal{F}$ at the particle location.
Therefore, the Lagrangian formulation derived above will enable rates calculation depending on $\mathcal{F}$ for the sake of the kinetic modeling.

To summerize, the novel aspects of the present paper are:

1.	A generalization of the "Furry picture" of the field theory under consideration. Accordingly, the massive-like dispersion relation obeyed by the laser photons emerges naturally from the formalism (analogously to the Higgs mechanism).

2.	The dispersion relation of a strong-field wave is found self consistently with the quantum solution of the Dirac / Klein Gordon equation.

3.	 Expression for the current of a single particle was obtained. With this formula at hand, we were able to solve the dispersion equation for a special case of a cold matter and demonstrated explicitly that in this case the photon effective mass is not affected by the spin, i.e. it is identical for a fermion and for a scalar.

4.	The Lagrangian we have written allows for cross sections calculation of QED processes, such as non-linear Compton and non-linear Breit-Wheeler, in the presence of a strong electromagnetic field in a medium. These processes are crucial for the creation of a QED cascade.

\section{Appendix A}
Let us calculate explicitly the current density of a fermion whose wavefunction is given by (\ref{eq:av psi_f2}).
\begin{multline}
j_{\mu} =\bar{u}_p \Bigl [  1+ c_k{\not}k + c_A {\not}A^{cl} + c_{kA} {\not}A^{cl}  {\not}k         \Bigr ]  \gamma_{\mu} \\
\Bigl [  1+ c_k{\not}k + c_A {\not}A^{cl} + c_{kA} {\not}A^{cl}  {\not}k         \Bigr ]u_p
\label{eq:av jmu_calc}
\end{multline}
The right hand side of Eq. (\ref{eq:av jmu_calc}) contains 16 terms
\begin{multline}
j_{\mu} =\bar{u}_p \Bigl(   [11]_{\mu} + [22]_{\mu} + [33]_{\mu} + [44]_{\mu} + \\  [12]_{\mu} + [13]_{\mu} + [14]_{\mu} + [23]_{\mu} + [24]_{\mu} + [34]_{\mu} \Bigr) u_p
\label{eq:av jmu_calc2}
\end{multline}
where $[ij] \equiv (\mbox{term }i) \times (\mbox{term }j) +   (\mbox{term }j) \times (\mbox{term }i)$. 

In the derivation below, the subsequent Dirac algebra identities are extensively useful
\begin{equation}
{\not}g{\not}f = 2g \cdot f - {\not}f{\not}g
\end{equation}
\begin{equation}
\gamma_{\mu}{\not}g = 2g_{\mu} -{\not}g \gamma_{\mu}
\end{equation}
where $g_{\mu}$ and $f_{\mu}$ are general 4-vectors.

In the following we evaluate the terms appearing in (\ref{eq:av jmu_calc2}). 
\begin{equation}
[11]_{\mu} = \gamma_{\mu}
\label{eq:av jmu_calc3}
\end{equation}
\begin{equation}
[22]_{\mu} = c_A^2 {\not}A^{cl}  \gamma_{\mu} {\not}A^{cl} = 2 {\not}A^{cl} A_{\mu}^{cl} + a^2 \gamma_{\mu}
\end{equation}
\begin{equation}
[33]_{\mu} = c_k^2 {\not}k  \gamma_{\mu} {\not}k = 2 {\not}k k_{\mu} - m_{ph}^2 \gamma_{\mu}
\end{equation}
\begin{multline}
[44]_{\mu} = c_{kA}^2  {\not}A^{cl} {\not}k \gamma_{\mu}{\not}k {\not}A^{cl} = 2 a^2 {\not}k k_{\mu}-   \\  2 m_{ph}^2 {\not}A^{cl} A_{\mu}^{cl} - m_{ph}^2 a^2 \gamma_{\mu}
\end{multline}
where we have used the fact $k \cdot A^{cl} = 0.$
\begin{equation}
[12]_{\mu} = c_A \left(  \gamma_{\mu} {\not}A^{cl} +   {\not}A\gamma_{\mu}     \right) = 2 c_A^{cl} A_{\mu}^{cl}
\end{equation}
\begin{equation}
[13]_{\mu} = c_k \left(  \gamma_{\mu} {\not}k +   {\not}k\gamma_{\mu}     \right) = 2 c_k k_{\mu}
\end{equation}
\begin{multline}
[14]_{\mu} = c_{kA} \left(    {\not}A^{cl} {\not}k \gamma_{\mu}   +  \gamma_{\mu}  {\not}k {\not}A^{cl}    \right) =  \\ 2 c_{kA} \left( {\not}A^{cl} k_{\mu} - {\not}k A_{\mu}^{cl}  \right)
\end{multline}
\begin{multline}
[23]_{\mu} = c_k c_A \left(   {\not}k\gamma_{\mu}{\not}A^{cl}  +  {\not}A^{cl} \gamma_{\mu}{\not}k   \right) = \\  2c_k c_A \left( {\not}A^{cl} k_{\mu}  +  {\not}k A_{\mu}^{cl}      \right)
\end{multline}
\begin{multline}
[24]_{\mu} = c_A c_{kA} \left(  {\not}A^{cl} \gamma_{\mu}{\not}k {\not}A^{cl}   + {\not}A^{cl} {\not}k \gamma_{\mu}{\not}A^{cl}     \right) =  \\ - 2 c_A c_{kA}  a^2 k_{\mu}     
\end{multline}
\begin{multline}
[34]_{\mu} = c_k c_{kA} \left(  {\not}k \gamma_{\mu}{\not}k {\not}A^{cl}   + {\not}A^{cl} {\not}k \gamma_{\mu}{\not}k     \right) = \\ -2 c_k c_{kA}  m_{ph}^2 A_{\mu}^{cl}    
\label{eq:av jmu_calc4}
\end{multline}
Now we substitute Eqs. (\ref{eq:av jmu_calc3} - \ref{eq:av jmu_calc4}) into Eq. (\ref{eq:av jmu_calc2}) and use $p \cdot A^{cl} = 0$ as well as the identity
\begin{equation}
\bar{u_p} \gamma_{\mu} u_p = \frac{\bar{u_p}  u_p}{m} p_{\mu}
\end{equation}
The final results reads
\begin{equation}
j_{\mu} =  \bar{u}_p u_p \left(  p_{\mu} j^p -   e A_{\mu}  j^A   -   k_{\mu} j^k   \right)
\end{equation}
where
\begin{equation}
j^A  \equiv -\frac{2}{e} \left[  \frac{(k \cdot p)}{m} \left(c_Ac_k - c_{kA} \right)  + c_A - c_k c_{kA}m_{ph}^2  \right]
\end{equation}
\begin{equation}
j^p  \equiv  \frac{ 1+a^2(c_A^2-m_{ph}^2c_{kA}^2) - c_k^2 m_{ph}^2 }{m}
\end{equation}
\begin{equation}
j^k  \equiv  -2 \left[  \frac{(k \cdot p)}{m} \left( c_k^2  +  a^2 c_{kA}^2 \right)  - a^2c_A c_{kA} + c_k  \right]
\end{equation}

\section{Appendix B}
We would like to outline the proof of the following algebraic identities 
\begin{equation}
j^p = j^A
\label{eq:av identity1}
\end{equation}
\begin{equation}
j^k = \nu j^p
\label{eq:av identity2}
\end{equation}
For this purpose we define the dimensionless variables
\begin{equation}
x \equiv \frac{ea}{m}
\end{equation}
\begin{equation}
y \equiv \frac{m_{ph}}{m}
\end{equation}
\begin{equation}
z \equiv \frac{k \cdot p}{m^2}
\end{equation}
\begin{equation}
w \equiv \sqrt{1+\left(  \frac{xy}{z} \right)^2}
\end{equation}
Now we write down the explicit expressions of $j^p, j^A, j^k$ in terms of $x,y,z,w$.
\begin{multline}
mj^p = 1+\frac{x^2}{16} \left(  1 - \frac{1}{w}  \right)^2 - y^2 \frac{x^2}{4 \left( z^2 +xy \right) } \\ - \frac{y^2}{4} \left[ \frac{z}{y^2} (1-w) +\frac{x^2}{2zw}  \right]^2
\end{multline}
\begin{multline}
-\frac{1}{2}mj^k = z \left\{ \frac{x^2}{4 \left(z^2 + xy \right)}  +\frac{1}{4} \left[   \frac{z}{y^2} (1-w) +\frac{x^2}{2zw}   \right]^2    \right\} \\ - \frac{1}{2} \left[ \frac{z}{y^2} (1-w) +\frac{x^2}{2zw}   \right]  + \frac{x^2}{8zw} \left( 1 - \frac{1}{w}  \right)
\end{multline}
\begin{multline}
-\frac{1}{2}mj^A = z \left[  \frac{1}{8}  \left(  1 - \frac{1}{w}  \right)  \left(  \frac{z}{y^2} (1-w) +\frac{x^2}{2zw}  \right) -\frac{1}{2zw}  \right] \\  -   \frac{1}{4} \left(   1 - \frac{1}{w}   \right)   + \frac{y^2}{4zw}  \left[  \frac{z}{y^2} (1-w) +\frac{x^2}{2zw}  \right]
\end{multline}
The above expressions can be simplified by finding the lowest common denominator and using the definition of $w$. After tedious but straightforward algebra one arrives at the required identites (\ref{eq:av identity1},\ref{eq:av identity2}).

\end{document}